\documentclass[aps,pra,twocolumn]{revtex4-1} 

\usepackage{verbatim}
\usepackage{dsfont} 
\usepackage{amsmath}
\usepackage{amsthm} 
\usepackage{enumerate}
\usepackage{graphicx}
\usepackage{mathtools} 
\usepackage{bm} 
\usepackage{amssymb} 
\usepackage{hyperref} 
\hypersetup{
    colorlinks=true,
    linkcolor=blue,
    citecolor=blue,
    unicode=true,          
}

\usepackage{blindtext}

\newcommand{\norm}[1]{\left\lVert#1\right\rVert}
\DeclareMathOperator{\Tr}{Tr}

\usepackage{calc} 
\usepackage{accents}
\newcommand{\dbtilde}[1]{\accentset{\approx}{#1}}

\makeatletter
\newcommand{\oset}[3][0ex]{%
  \mathrel{\mathop{#3}\limits^{
    \vbox to#1{\kern-2\ex@
    \hbox{$\scriptstyle#2$}\vss}}}}
\makeatother

\usepackage{diagbox}
\begin{document}


\title{Multi-object operational tasks for convex quantum resource theories} 


\author{Andr\'es F. Ducuara$^{1,2}$, Patryk Lipka-Bartosik$^{3}$ and Paul Skrzypczyk$^{3}$} 

\address{$^{1}$Quantum Engineering Centre for Doctoral Training, \\H. H. Wills Physics Laboratory and Department of Electrical \& Electronic Engineering, University of Bristol, BS8 1FD, UK}

\address{$^{2}$Quantum Engineering Technology Labs, H. H. Wills Physics Laboratory and \\
Department of Electrical \& Electronic Engineering, University of Bristol, BS8 1FD, UK.}

\address{$^{3}$H.H. Wills Physics Laboratory, University of Bristol, Tyndall Avenue, Bristol, BS8 1TL, United Kingdom}


\date{\today}

\begin{abstract}
The prevalent modus operandi within the framework of quantum resource theories has been to characterise and harness the resources within single objects, in what we can call \emph{single-object} quantum resource theories. One can wonder however, whether the resources contained within multiple different types of objects, now in a \emph{multi-object} quantum resource theory, can simultaneously be exploited for the benefit of an operational task. In this work, we introduce examples of such multi-object operational tasks in the form of subchannel discrimination and subchannel exclusion games, in which the player harnesses the resources contained within a state-measurement pair. We prove that for any state-measurement pair in which either of them is resourceful, there exist discrimination and exclusion games for which such a pair outperforms any possible free state-measurement pair. These results hold for arbitrary convex resources of states, and arbitrary convex resources of measurements for which classical post-processing is a free operation. Furthermore, we prove that the advantage in these multi-object operational tasks is determined, in a multiplicative manner, by the resource quantifiers of: \emph{generalised robustness of resource} of both state and measurement for discrimination games and \emph{weight of resource} of both state and measurement for exclusion games. 
\end{abstract} 


\maketitle

\section{Introduction}

The framework of quantum resource theories (QRTs) \cite{RT_review} has proven to be a successful approach to quantum information theory. Broadly speaking, it aims at identifying, characterising, and utilising quantum phenomena as a \emph{resource} for fuelling quantum information processing protocols for the development of quantum technologies. A QRT is specified by first defining the \emph{objects} of the theory, followed by a \emph{property} of these objects to be regarded as a resource. The choice of a particular property as a resource is then justified by specifying instances, usually in the form of operational tasks, in which the presence of such resourceful object provides an advantage over all resourceless (free) objects. There is a plethora of objects in quantum theory whose properties are deemed as resources, namely: states \cite{QRTE1,QRTE2}, measurements \cite{RT_measurements0, RT_measurements1, RT_measurements2}, behaviours or boxes \cite{RT_nonlocality, RT_noncontextuality}, steering assemblages \cite{RT_steering}, teleportation assemblages \cite{RoT} and channels \cite{RT_channels1, RT_channels2, Plenio, Pirandola}, amongst many others \cite{RT_review, Schmid1, Schmid2, Schmid3, Schmid4}. Arguably, the most studied QRTs are the ones for \emph{states} and \emph{measurements}. QRTs of states address resources like: entanglement \cite{QRTE2}, coherence \cite{RT_coherence}, asymmetry \cite{RT_coherence}, superposition \cite{RT_superposition}, purity \cite{RoP}, magic \cite{RT_magic}, amongst many others \cite{RT_nongaussianity, RT_nonmarkovianity1, RT_nonmarkovianity2, namit, RT_thermodynamics, RT_RF}. QRTs of measurements on the other hand, address resources like: entanglement \cite{RT_measurements1}, coherence \cite{RT_measurements1}, informativeness \cite{RoM} and non-projective simulability \cite{RT_PS}.

Having specified a set of objects and one of their properties to be treated as a resource, it is of interest to quantitatively specify the \emph{amount} of resource contained within a given object. This can be accomplished by introducing appropriate measures known as \emph{resource quantifiers} \cite{RT_review}. Two well-known families of these measures are the so-called \emph{robustness-based} \cite{RoE, RoNL_RoS_RoI, RoS, RoA, RoC, RoM, RoNL_RoS_RoI, RoT, RoT2, RT_magic, citeme1, citeme2} and \emph{weight-based} \cite{WoE,EPR2,WoS,RoNL_RoS_RoI,WoAC} resource quantifiers. These quantifiers have found applications in several scenarios, for instance, at characterising the advantage that a resourceful object offers, when compared to all resourceless objects, in specific \emph{operational tasks}. There are two broad families of such results addressing this \emph{quantifier-task correspondence}, the so-called robustness-discrimination \cite{RoS, RoI_task, RoI_Channels, RoA, RoM, RT2} and weight-exclusion \cite{QRT1, Uola, QRT2} correspondences.

One common feature amongst all of these results is that they address \emph{single-object} operational tasks, meaning that a \emph{single} object is thought of as the resourceful object, and the associated tasks are then exploiting the resource contained within such an individual object. One then can wonder, about the possibility of having operational tasks harnessing \emph{two or more} different resources out of two, in principle different, objects. We refer to these tasks as \emph{multi-object} tasks, and we can intuitively approach them from the following two general levels. In a first instance, one can consider a single QRT with two different resources, in which case it is natural to make the distinction of the resources being either: disjoint, intersecting or nested \cite{QRT_FL}. The case of QRTs of states with \emph{disjoint} resources has been explored in the context of a first law for general QRTs \cite{QRT_FL}, this, inspired by results from the thermodynamics of multiple conserved quantities \cite{MC1,MC2,MC3}. In a second instance however, one can also consider a multi-object scenario in which a first QRT of certain objects with an arbitrary resource is being specified, followed by a second QRT with \emph{different} objects with their respective arbitrary resource. In this work we address this latter case by considering a multi-object scenario with two objects, one being a \emph{state} and a second one being a \emph{measurement} and therefore, the composite object of interest is now a state-measurement \emph{pair}.

In this work we address composite QRTs made of convex QRTs of states with arbitrary resources and convex QRTs of measurements with arbitrary resources for which the operation of classical post-processing (CPP) is a \emph{free operation}, meaning that it takes free (resourceless) measurements into free measurements. Taking into account that CPP is a free operation for many important resources for measurements like: entanglement, coherence and informativeness, the results found in this work naturally apply to all of these instances. Explicitly, we introduce \emph{multi-object} operational tasks in the form of subchannel discrimination and subchannel exclusion games in which, a state-measurement pair is being deemed as the composite object of the theory, as opposed to the state (or the measurement) alone. Interestingly, we have found that \emph{any} resourceful state-measurement pair offers an advantage, over \emph{all} possible free pairs, when performing at particular multi-object tasks. Furthermore, we have found that this advantage can be \emph{quantified}, in a multiplicative manner, by the amount of resource contained within each object, here measured by the resource quantifiers of \emph{generalised robustness} and \emph{weight}, for discrimination and exclusion games respectively. Moreover, these quantifiers also find operational significance in an multi-object encoding-decoding communication task involving the state-measurement pair. We believe that the results found in this work open the door for the exploration of multi-object operational tasks in general convex QRTs with different objects beyond states and measurements.

\section{Composite convex quantum resource theories and multi-object operational tasks}

We start by addressing convex QRTs of states and measurements with arbitrary resources.

\textbf{Definition 1:} (Composite convex QRTs of states and measurements) Consider the set of quantum states in a complex finite-dimensional Hilbert space. A quantum state is an operator $\rho$ satisfying $\rho\geq 0$ and $\Tr(\rho)=1$. We now consider a property of these states defining a closed convex set which we will call the set of free states and denote as ${\rm F}$. We say a state $\rho \in {\rm F}$ is a \emph{free (resourceless)} state and it is \emph{resourceful} otherwise. We also consider the set of quantum measurements, or positive-operator valued measures (POVMs) in the same complex finite-dimensional Hilbert space. A POVM is a collection of operators $\mathbb{M}=\{M_a\}$, $a\in \{1,...,o\}$ with $M_a\geq 0$, $\forall a$ and $\sum_{a=1}^{o} M_a = \mathds{1}$. Similarly, we consider a property of measurements defining a closed convex set of free measurements and denote it as $\mathbb{F}$. We say a POVM $\mathbb{M} \in \mathbb{F}$ is a \emph{free (resourceless)} measurement and it is \emph{resourceful} otherwise. We say that a state-measurement pair $(\rho,\mathbb{M})$ is: \emph{fully free} when both state and measurement are free, \emph{partially resourceful} when either is resourceful, and \emph{fully resourceful} when both are resourceful.

We now address an operation for measurements.

\textbf{Definition 2:} (Classical post-processing (CPP)) We say that a measurement $\mathbb{N}=\{N_x\}$, $x\in \{1,...,k\}$ is \emph{simulable by} the measurement $\mathbb{M}=\{M_a\}$, $a\in \{1,...,o\}$ when there exists a conditional probability distribution $\{q(x|a)\}$ such that $N_x=\sum_{a=1}^{o} q(x|a)M_a$, $\forall x\in \{1,...,k\}$ \cite{simulability}. One can check that the simulability of measurements defines a partial order for the set of measurements and therefore we use the notation $\mathbb{N} \preceq \mathbb{M}$, meaning that $\mathbb{N}$ is simulable by $\mathbb{M}$. We refer to this property as simulability of measurements or classical post-processing (CPP).

We can check that CPP is a free operation for QRTs of measurements with the resources of: entanglement, coherence, informativeness and non-projective simulability. We will then be addressing, from now on, convex QRTs of measurements for which CPP is a free operation. We now introduce multi-object operational tasks which are meant to be played with state-measurement pairs.

\textbf{Definition 3:} (Multi-object subchannel discrimination/exclusion games) Consider a player with access to a state-measurement pair $(\rho, \mathds{M})$. The player sends the state $\rho$ to the referee who is in possession of a collection of subchannels $\Psi=\{\Psi_x\}$, $x\in\{1,...,k\}$. The subchannels $\{\Psi_x\}$ are completely-positive (CP) trace-nonincreasing linear maps, such that $\sum_x \Psi_x$ forms a completely-positive trace-preserving (CPTP) linear map. The referee promises to apply one of these subchannels on the state $\rho$ and the transformed state is then sent back to the player. The player then effectively has access to the ensemble $\mathcal{E}^\rho_{\Psi}=\{\rho_x, p(x)\}$ with $p(x)=\Tr[\Psi_x(\rho)]$, $\rho_x=\Psi_x(\rho)/p(x)$. In a \emph{subchannel discrimination} game, the goal is for the player to output a guess $g \in \{1,...,k\}$ for the subchannel that was applied, the player succeeds at the game if $g = x$ and fails when $g\neq x$. In a \emph{subchannel exclusion} game on the other hand, the goal is for the player to output a guess $g \in \{1,...,k\}$ for a subchannel that was \textit{not} applied, that is, the player succeeds at the game if $g \neq x$ and fails when $g=x$. In order to generate a guess, the player proceeds to implement the measurement $\mathbb{M}=\{M_a\}$ on the received state and classically post-process the measurement outcome $a$ to produce an output guess $g$, according to a probability distribution $\{p(g|a)\}$, for playing either a discrimination or an exclusion game. The probability of success at subchannel discrimination and the probability of error at subchannel exclusion are given by:
\begin{align}
    P^{\rm D}_{\rm succ}(\Psi,\rho,\mathbb{M})
    &=
    \max_{\{p(g|a)\}} 
    \sum_{x,a,g}
    \delta_{x,g} \, p(g|a)\, p(a|x) \, p(x),
    \label{eq:QScD}\\
    P^{\rm E}_{\rm err}(\Psi,\rho,\mathbb{M})
    &=
    \min_{\{p(g|a)\}} 
    \sum_{x,a,g}
    \delta_{x,g} \, p(g|a)\, p(a|x) \, p(x),
    \label{eq:QScE}
\end{align}
with $p(a|x)=\Tr[M_a\rho_x]$ and the maximisation (minimisation) over all classical post-processings of the measurements outputs $p(g|a)$. A subchannel discrimination/exclusion game is specified by the collection of subchannels $\Psi=\{\Psi_x\}$. 

A key point to remark, is that the object of interest is now the state-measurement pair $(\rho,\mathbb{M})$, as opposed to the state (or measurement) alone. We now proceed to establish a first result comparing the performance of a \emph{fully resourceful} state-measurement pair against all \emph{fully free} pairs when addressing a particular game.

\section{Any fully resourceful state-measurement pair is useful for multi-object subchannel discrimination/exclusion}

\textbf{Result 1:} Consider a convex QRT of states with an arbitrary resource and a convex QRT of measurements with an arbitrary resource for which CPP is a free operation. Given a \emph{fully resourceful} state-measurement pair $(\rho,\mathbb{M})$, meaning  that we have both a resourceful state $\rho \notin {\rm F}$ and a resourceful measurement $\mathbb{M}\notin\mathbb{F}$, then, there exist subchannel games $\Psi_D^{(\rho,\mathbb{M})}$ and $\Psi_E^{(\rho,\mathbb{M})}$ such that:
\begin{align}
    \max_{\sigma \in {\rm F}}
    \max_{\mathbb{N} \in \mathbb{F}}
    P^{\rm D}_{\rm succ}
    (
    \Psi_D^{(\rho,\mathbb{M})},\sigma,\mathbb{N}
    )
    <
    P^{\rm D}_{\rm succ}
    (
    \Psi_D^{(\rho,\mathbb{M})},\rho,\mathbb{M}
    )
    \label{eq:result1D},\\
    P^{\rm E}_{\rm err}(\Psi_E^{(\rho,\mathbb{M})},\rho, \mathbb{M})
    <
    \min_{\sigma \in {\rm F}}
    \min_{\mathbb{N} \in \mathbb{F}}
    P^{\rm E}_{\rm err}(\Psi_E^{(\rho,\mathbb{M})},\sigma,\mathbb{N}).
    \label{eq:result1E}
\end{align}
These two \emph{strict} inequalities mean that the state-measurement pair $(\rho,\mathbb{M})$ provides \emph{strictly} larger (smaller) advantage (error) than \emph{all} fully free state-measurement pairs, when playing the subchannel discrimination (exclusion) game specified by $\Psi_D^{(\rho,\mathbb{M})}$ ($\Psi_E^{(\rho,\mathbb{M})}$). 

The proof of this result relies on the hyperplane separation theorem \cite{CA_book} as well as on a trick first used in the context of quantum steering \cite{RoS}, for ``completing'' a set of subchannels, from which one can extract suitable operators in order to construct the tailored subchannel games $\Psi_D^{(\rho,\mathbb{M})}$ and $\Psi_E^{(\rho,\mathbb{M})}$, for which playing with the pair $(\rho,\mathbb{M})$ is optimal. The full proof of this result is in Appendix A. We now would like to \emph{quantify} this advantage by specifying how large this gap can be. In order to do this, we need to define a suitable resource quantifier for the composite objects of state-measurement pairs. A natural starting point is to quantify the amount of resource contained within the individual objects of interest, states and measurements. 

\section{Resource quantifiers and multi-object games}

We now address resource quantifiers for convex QRTs of states and measurements with arbitrary resources.

\textbf{Definition 4:} (Generalised robustness and weight for states and measurements) Consider a convex QRT of states with an arbitrary resource and a convex QRT of measurements with an arbitrary resource. The generalised robustness and the weight of resource of a state and a measurement are given by:
\begin{align}
    {\rm R_F}\left(\rho\right)
    &=
    {\scriptsize
    \begin{matrix}
    \text{\small \rm min}\\
    r \geq 0\\
    \sigma \in {\rm F} \\
    \rho^G \\
    \end{matrix}
    }
    \left\{ 
    \rule{0cm}{0.6cm} r\,\bigg| \, \rho+r\rho^G=(1+r)\sigma
    \right\},
    \label{eq:RoRs}\\
    {\rm R_\mathbb{F}}\left(\mathbb{M}\right)
    &=
    {\scriptsize
    \begin{matrix}
    \text{\small \rm min}\\
    r \geq 0\\
    \mathbb{N} \in \mathbb{F} \\
    \mathbb{M}^G \\
    \end{matrix}
    }
    \left\{ 
    \rule{0cm}{0.6cm} r\,\bigg| \, M_a+rM^G_a=(1+r)N_a
    \right\},
    \label{eq:RoRm}\\
    {\rm W_F}\left(\rho\right)
    &=
    {\scriptsize
    \begin{matrix}
    \text{\small \rm min}\\
    w \geq 0\\
    \sigma \in {\rm F} \\
    \rho^G \\
    \end{matrix}
    }
    \left\{ 
    \rule{0cm}{0.6cm} w\,\bigg| \, \rho=w\rho^G+(1-w)\sigma
    \right\},
    \label{eq:WoRs}\\
    {\rm W_\mathbb{F}}\left(\mathbb{M}\right)
    &=
    {\scriptsize
    \begin{matrix}
    \text{\small \rm min}\\
    w \geq 0\\
    \mathbb{N} \in \mathbb{F} \\
    \mathbb{M}^G \\
    \end{matrix}
    }
    \left\{ 
    \rule{0cm}{0.6cm} w\,\bigg| \, M_a=wM^G_a+(1-w)N_a
    \right\}.
    \label{eq:WoRm}
\end{align}
The generalised robustness quantifies the minimum amount of a general state $\rho^G$ (measurement $\mathbb{M}^G$) that has to be added to $\rho$ ($\mathbb{M}$) such that we get a free state $\sigma$ (measurement $\mathbb{N}$). The weight on the other hand, quantifies the minimum amount of a general state $\rho^G$ (measurement $\mathbb{M}^G$) that has to be used for recovering the state $\rho$ (measurement $\mathbb{M}$). 

One now would like to introduce a quantifier for the composite object $({\rho,\mathbb{M}})$. It turns out however, that it is enough to quantify the resources contained within the individual objects, as we will see in what follows. We now establish a connection between robustness-based (weight-based) resource quantifiers for states and measurements and multi-object subchannel discrimination (exclusion) games.

\textbf{Result 2:} Consider a convex QRT of states with an arbitrary resource and a convex QRT of measurements with an arbitrary resource for which CPP is a free operation. Given \emph{any} state-measurement pair $(\rho, \mathbb{M})$ we have:
\begin{align}
    \hspace{-0.1cm}
    \max_{\rm \Psi} \frac{
    P^{\rm D}_{\rm succ}(\Psi,\rho,\mathbb{M})
    }{
    \displaystyle
    \max_{
    \substack{
    \sigma \in {\rm F}\\
    \mathbb{N} \in \mathbb{F}
    }
    }
    P^{\rm D}_{\rm succ}(\Psi,\sigma,\mathbb{N})
    }
    = \hspace{-0.1cm}
    \Big [1+{\rm R_F}(\rho)\Big]
    \Big[1+{\rm R_\mathbb{F}}(\mathbb{M}) \Big],
    \label{eq:result2D}
\end{align}
\vspace{-0.5cm}
\begin{align}
    \hspace{-0.11cm}\min_{\rm \Psi} \frac{
    P^{\rm E}_{\rm err}(\Psi,\mathbb{M},\rho)
    }{
    \displaystyle
    \min_{
    \substack{
    \sigma \in {\rm F}\\
    \mathbb{N} \in \mathbb{F}
    }
    }
    P^{\rm E}_{\rm err}(\Psi,\sigma,\mathbb{N})
    }= 
    \hspace{-0.1cm}
    \Big [1-{\rm W_F}(\rho)\Big]
    \Big[1-{\rm W_\mathbb{F}}(\mathbb{M}) \Big],
    \label{eq:result2E}
\end{align}
with the maximisation (minimisation) over all subchannel games.

The full proof of this result is in Appendix B. The first thing we can notice is, that by considering a \emph{fully resourceful} state-measurement pair $(\rho,\mathbb{M})$, one recovers the strict inequalities in \eqref{eq:result1D} and \eqref{eq:result1E}. Additionally, we can also see that by considering now a \emph{partially resourceful} pair $(\rho,\mathbb{M})$, meaning that either the state or the measurement is resourceful, we still get an advantage. This may seem counter-intuitive at first sight, as using a resourceless measurement should not allow the player to obtain any advantage, even with the most resourceful state. However, as we explicitly show in Appendix B, there still exists a game which allows the player to utilise the advantage arising in such a partially-resourceful scenario. The resolution to this apparent paradox is based on the crucial difference between channel and subchannel discrimination/exclusion tasks. In particular, in a subchannel discrimination/exclusion game, a resourceful state has the additional ability to ``influence'' the ensemble of states from which the player needs to discriminate/exclude, this, since $\mathcal{E}^\rho_{\Psi}=\{\rho_x, p(x)\}$ with $p(x)=\Tr[\Psi_x(\rho)]$, $\rho_x=\Psi_x(\rho)/p(x)$ and therefore, this lead to suitable ensembles, even for resourceless measurements. Finally, for a fixed \emph{fully free} pair, there exists a game for which the pair is still optimal amongst all free pairs. Therefore, the ratios considered in Result 2 are comparing the performance of \emph{any} pair against \emph{all} fully free pairs. 

It is illustrative to compare these results with their \emph{single-object} counterparts \cite{RT1, QRT2}. When considering subchannel games being played with a state alone, and allowing maximisations over arbitrary measurements, the advantage becomes $[1+{\rm R_F}(\rho)]$ \cite{RT1}. In the multi-object scenario considered here however, we get $[1+{\rm R_F}(\rho)][1+{\rm R}_\mathbb{M}(\mathbb{M})]$ instead, which can get to be larger, whenever $\mathbb{M}$ is resourceful. A similar analysis can be made for the weight-exclusion case \cite{QRT2}. This increment can be conceptually understood by the fact that we are now addressing a composite object and therefore, it is natural that each object contributes to the overall advantage. Nevertheless, it is still surprising that the advantage can be quantified in this elegant multiplicative manner. 

It is also interesting to note that this result applies to convex QRTs of states with \emph{arbitrary resources} and convex QRTs of measurements with \emph{arbitrary resources} for which CPP is a free operation and therefore it covers, as particular instances, several important resources for both states and measurements. It would be interesting to explore whether these results still hold when CPP is dropped or, on the other hand, if a counterexample can be found. We leave this however for future research. We now address multi-object single-shot information-theoretic quantities in the context of an encoding-decoding communication task.

\section{Single-shot information theory}

Consider a state-measurement pair $(\rho,\mathbb{M})$ and the following  communication task. A random variable $X$ is going to be encoded, with the help of the state $\rho$ and an ensemble of channels $\Lambda=\{\Lambda_x, p(x)\}$, in the ensemble of states $\mathcal{E}^{(\Lambda,\rho)}=\{\sigma^{(\Lambda,\rho)}_x,p(x)\}$ with $\sigma^{(\Lambda,\rho)}_x=\Lambda_x(\rho)$. We refer to the classical random variable $X$ encoded in such a way as $X_{\Lambda,\rho}$. We then consider a decoding scheme using the measurement $\mathbb{M}=\{M_g\}$ with its outcomes representing a (guess) classical random variable $G$. Similarly, we refer to such a decoded variable as $G_\mathbb{M}$. We then have that this encoding-decoding scheme depends on the state-measurement pair $(\rho,\mathbb{M})$. A well studied figure of merit for communication tasks is the so-called \emph{accessible information} \cite{Wilde_book}. Additionally, it has recently been introduced a complementary figure of merit which has been coined the \emph{excludible information}, for its natural connection to exclusion tasks \cite{QRT1, QRT2}. These quantities depend on the plus (minus) infinity mutual information (respectively), which are given by:
\begin{align*}
    I_{\pm \infty}(X_{\Lambda,\rho}:G_\mathbb{M})=
    \pm
    \Big[
    H_{\pm \infty}(X_{\Lambda,\rho})-H_{\pm \infty}(X_{\Lambda,\rho}|G_\mathbb{M})
    \Big],
\end{align*}
with the order plus and minus infinity entropies $H_{+\infty}(X_{\Lambda,\rho})=-{\rm log}\, \left\{\max_{x} p(x)\right\}$ and $H_{-\infty}(X_{\Lambda,\rho})=-{\rm log}\, \left\{ 
\min_{x} p(x) \right\}$, the order plus and minus infinity conditional entropies $H_{+\infty}(X_{\Lambda,\rho}|G_\mathbb{M})=-{\rm log} \left\{
\sum_g \max_x \, p(x,g) \right\}$ and $H_{-\infty}(X_{\Lambda,\rho}|G_\mathbb{M})=-{\rm log} \left\{
\sum_g \min_x \,p(x,g) \right\}$, with $p(x,g)=p(g|x)p(x)$ and $p(g|x)=\Tr[M_g\Lambda_x(\rho)]$. The $\pm \infty$ mutual information quantifies the amount of the respective type of information (accessible or excludible) that can be conveyed by the state-measurement pair and the ensemble of channels at play. These measures are usually functions of the channel but we consider them here as functions of the $(\rho,\mathbb{M})$ pair instead. We now address these quantities for a state-measurement pair in comparison to \emph{all} fully free pairs.

\textbf{Result 3:} Consider a state-measurement pair $(\rho,\mathbb{M})$. The maximum gap between the plus (minus) infinity mutual information between this pair and \emph{all} fully free state-measurement pairs is upper bounded as:
\begin{multline}
    \max_{\Lambda} 
    \left\{
    I_{+\infty}(X_{\Lambda,\rho} \colon G_{\mathbb{M}})
    -
    \max_{\sigma\in {\rm F}}
    \max_{\mathbb{N}\in \mathbb{F}}
    I_{+\infty}(X_{\Lambda,\sigma} \colon G_{\mathbb{N}})
    \right\}\\
    \leq
    \log \Big[1+{\rm R_F}(\rho)\Big]
    +
    \log \Big[1+{\rm R_\mathbb{F}}(\mathbb{M})\Big],
    \label{eq:result3R}
\end{multline}
\vspace{-1cm}
\begin{multline}
    \max_{\Lambda} 
    \left\{
    I_{-\infty}(X_{\Lambda,\rho} \colon G_{\mathbb{M}})
    -
    \max_{\sigma \in {\rm F}}
    \max_{\mathbb{N} \in \mathbb{F}}
    I_{-\infty}(X_{\Lambda,\sigma} \colon G_{\mathbb{N}})
    \right\}\\
    \leq
    -\log \Big[1-{\rm W_F}(\rho)\Big]
    -\log \Big[1-{\rm W_\mathbb{F}}(\mathbb{M})\Big],
    \label{eq:result3W}
\end{multline}
with the maximisation over all ensembles of channels.

The full proof of this result is in Appendix C. This result means that the resource quantifiers are placing upper bounds for these quantities. It would be interesting to see whether they can be saturated.

\section{Conclusions}

In this work we have introduced \emph{multi-object} operational tasks in which the composite objects of interest are state-measurement pairs. The results found in this article hold for convex QRTs of states with arbitrary resources and convex QRTs of measurements for which CPP is a free operation. In particular, we have shown that \emph{any} resourceful pair is useful for multi-object subchannel discrimination and exclusion games, when compared to the best possible strategy using \emph{fully free} state-measurement pairs. Furthermore, we have found that this advantage can be quantified, in a \emph{multiplicative manner}, by the quantifiers of generalised robustness and weight of the state and the measurement, for discrimination and exclusion respectively. These results also provide support, now  in the multi-object regime, to the conjecture made in \cite{QRT1}, about the existence of a weight-exclusion correspondence whenever there is a robustness-discrimination one. We have also introduced a communication task in which the log-robustness and the log-weight place upper bounds for information-theoretic quantities.  

We believe that this work opens the door for exploring \emph{multi-object} operational tasks in general QRTs of \emph{arbitrary composite objects} with \emph{arbitrary resources}, beyond those of states and measurements, as well as tasks for pairs of the same type of objects but exploiting different resources, and whether the distinction between the resources being disjoint, intersecting and nested plays any major role.

\section*{Acknowledgements}

We would like to thank Tom Purves for insightful discussions. AFD acknowledges support from COLCIENCIAS 756-2016. PLB acknowledges support from the UK EPSRC (grant no. EP/R00644X/1). PS acknowledges support from a Royal Society URF (UHQT).

\bibliographystyle{apsrev4-1}
\bibliography{bibliography.bib}

\appendix
\section{Proof of Result 1}

In order to prove Result 1, we start by rewriting the figures of merit in a more compact form, we then extract some useful operators using the hyperplane separation theorem and define a particular classical post-processing (CPP) operation. With this in place, we proceed to address the discrimination case followed by the exclusion case.

\subsection{Rewriting the figures of merit}

We start by rewriting the probability of success (error) in multi-object discrimination (exclusion) games in a more compact form. Given a multi-object discrimination game $\Psi=\{\Psi_x(\cdot)\}$ and a state-measurement pair $(\rho,\mathbb{M})$, the probability of success can be written as:
\begin{align*}
    &P_{\rm succ}^{\rm D}(\Psi,\rho,\mathbb{M})\\
    &= \max_{\{p(g|a)\}} 
    \sum_{x,a,g}
    \delta_{x,g} \, p(g|a)\, p(a|x) \, p(x),\\
    &= \max_{\{p(g|a)\}} 
    \sum_{x,a,g}
    \delta_{x,g} p(g|a) 
    \Tr\left[
    M_a \frac{\Psi_x(\rho)}{\Tr \left[\Psi_x(\rho) \right]}
    \right]
    p(x),\\
    &= \max_{\{p(g|a)\}} 
    \sum_{x,a,g}
    \delta_{x,g} p(g|a) \Tr[M_a \Psi_x(\rho)],\\
    &= \max_{\{p(g|a)\}} 
    \sum_{x} {\rm Tr}\left\{ \left[\sum_{a} \left(\sum_{g} p(g|a) \delta_{x,g}\right) M_a \right] \Psi_x(\rho) \right\},\\
    &=\max_{\{p(x|a)\}} 
    \sum_{x}
    \Tr \left\{\left[\sum_{a} p(x|a) M_a \right]\Psi_x(\rho) \right \},\\
    &=\max_{\mathbb{N} \preceq \mathbb{M}} 
    \sum_x \Tr\left[N_x\Psi_x(\rho)\right],
\end{align*}
where in the third line we used $p(x)=\Tr[\Psi_x(\rho)]$, and in the last line the maximisation is over all measurements simulable by $\mathds{M}$. Similarly, for the multi-object exclusion case we get:
\begin{align*}
    P_{\rm err}^{\rm E}(\Psi,\rho,\mathbb{M})
    =\min_{\mathbb{N} \preceq \mathbb{M}}
    \sum_x \Tr\left[N_x\Psi_x(\rho)\right].
\end{align*}

\subsection{Some useful operators}

Given any \emph{fully resourceful} state-measurement pair $(\rho,\mathbb{M})$, meaning that $\rho \notin {\rm F}$ and $\mathbb{M}=\{M_x\}\notin \mathbb{F}$, $x\in\{1,...,k\}$ and using the hyperplane separation theorem \cite{CA_book}, we have that there exist positive semidefinite operators $Z^\rho$ and $\{Z^\mathbb{M}_x\}$, $x\in\{1,...,k\}$ such that:
\begin{align}
    \Tr(Z^\rho \rho)>1,\hspace{0.5cm}
    \sum_x\Tr(Z^\mathbb{M}_x M_x)>1,
    \label{eq:condition1}\\
    \Tr(Z^\rho \sigma)\leq1, \hspace{0.5cm}
    \sum_x\Tr(Z^\mathbb{M}_x N_x)\leq1,
    \label{eq:condition2}
\end{align}
$\forall \sigma \in {\rm F}, \mathbb{N} \in \mathbb{F}$. Similarly, there exist positive semidefinite operators $Y^\rho$ and $\{Y^\mathbb{M}_x\}$, $x\in\{1,...,k\}$ such that:
\begin{align}
    \Tr(Y^\rho \rho)<1,\hspace{0.5cm}
    \sum_x\Tr(Y^\mathbb{M}_x M_x)<1,
    \label{eq:condition3}\\
    \Tr(Y^\rho \sigma)\geq1, \hspace{0.5cm}
    \sum_x\Tr(Y^\mathbb{M}_x N_x)\geq1,
    \label{eq:condition4}
\end{align}
$\forall \sigma \in {\rm F}, \mathbb{N} \in \mathbb{F}$. These sets of operators are going to be useful when constructing the subchannel games for discrimination and exclusion.

\subsection{Particular CPP operation}

Given an arbitrary measurement $\mathbb{N}=\{N_a\}$ with $a\in\{1,...,k+n\}$, $n$ and $k$ integers, we then construct a measurement $ \mathbb{\tilde N}=\{\tilde N_x\}$ with $k$ elements as:
\begin{align}
    \nonumber \tilde N_x& 
    \coloneqq 
    N_x, \hspace{1cm}
    x\in \{1,...,k-1\},\\
    \tilde N_k& 
    \coloneqq 
    N_k+\sum_{y=k+1}^{k+n}N_y. 
    \label{eq:Ntilde}
\end{align}
We can check that this is a well-defined measurement and that the operation taking $\mathbb{N}$ into $\mathbb{\tilde N}$ is a CPP operation on the initial measurement $\mathbb{N}$. This corresponds to a coarse graining of measurement outcomes, such that any outcome of $\mathds{N}$ greater or equal than $k$ is declared as outcome $k$.

\subsection{Discrimination case}

\textbf{Result 1A:} Consider a convex QRT of states with an arbitrary resource and a convex QRT of measurements with an arbitrary resource for which CPP is a free operation. Given any \emph{fully resourceful} state-measurement pair $(\rho,\mathbb{M})$, meaning that we have a resourceful state $\rho \notin {\rm F}$ and a resourceful measurement $\mathbb{M}\notin\mathbb{F}$, then, there exists a subchannel game $\Psi^{(\rho,\mathbb{M})}$ such that:
\begin{align}
    \hspace{-0.1cm}\max_{\sigma \in {\rm F}}
    \max_{\mathbb{N} \in \mathbb{F}}
    P^{\rm D}_{\rm succ}
    (
    \Psi^{(\rho,\mathbb{M})},\sigma,\mathbb{N}
    )
    <
    P^{\rm D}_{\rm succ}
    (
    \Psi^{(\rho,\mathbb{M})},\rho,\mathbb{M}
    ),
    \label{eq:result1DA}
\end{align}
with the left side being maximised over all possible free states and free measurements.

\begin{proof}
We start by considering a \emph{fully resourceful} state-measurement pair $(\rho,\mathbb{M})$. Using the hyperplane separation theorem \cite{CA_book}, there exist positive semidefinite operators $Z^\rho$ and $\{Z^\mathbb{M}_x\}$, $x\in\{1,...,k\}$ satisfying the conditions \eqref{eq:condition1} and \eqref{eq:condition2}. We now define the set of maps $\{\Phi^{(\rho,\mathbb{M})}_x (\cdot)\}$ such that for any state $\eta$ we have:
\begin{align*}
    \Phi^{(\rho,\mathbb{M})}_x(\eta)
    &\coloneqq
    \alpha^{(\rho,\mathbb{M})}
    \Tr(Z^\rho \eta)
    Z^\mathbb{M}_x,\\
    \alpha^{(\rho,\mathbb{M})}
    &\coloneqq 
    \frac{1}{
    \norm{Z^\rho}_1
    \Tr(Z^\mathbb{M})
    },\hspace{0.5cm}
    Z^\mathbb{M} \coloneqq
    \sum_{x=1}^{k} Z^\mathbb{M}_x,
\end{align*}
with $\norm{X}_1=\Tr(\sqrt{X^\dagger X})$ the trace norm. We are going to use the notation $\alpha =\alpha^{(\rho,\mathbb{M})}$. We can check that these maps are completely-positive and linear, and that they satisfy that $\forall \eta$:
\begin{align*}
    F(\eta) 
    \coloneqq 
    \Tr \left[
    \sum_{x=1}^{k} \Phi^{(\rho,\mathbb{M})}_x(\eta)
    \right]=
    \frac{\Tr(Z^\rho \eta)}{\norm{Z^\rho}_1}
    \leq 1.
\end{align*}
The inequality follows from the variational characterisation of the trace norm, establishing that $\norm{X}_1=\max_{-\mathds{1}\leq M\leq \mathds{1}}\{\Tr(XM)\}$ for any Hermitian operator $X$ \cite{Wilde_book}. We can also write $F(\eta)=\alpha \Tr(Z^\rho \eta) \Tr(Z^\mathbb{M})$. The set of maps $\{\Phi^{(\rho,\mathbb{M})}_x(\cdot)\}$ then add up to a completely positive trace-nonincreasing linear map. We can then \emph{complete} this set to be a set of subchannels by adding an extra subchannel $\Psi^{(\rho, \mathbb{M})}_{k+1}(\cdot) \coloneqq \Lambda(\cdot)- \Phi^{(\rho,\mathbb{M})}(\cdot)$, with $\Lambda$ being an arbitrary CPTP map such that it is greater or equal than zero (take the identity channel for instance). Therefore, with this construction we obtain a well-defined set of \emph{subchannels} with $k+1$ elements. We now proceed to define a family of sets of subchannels in the following manner. Given a state-measurement pair $(\rho,\mathbb{M})$, $\mathbb{M}=\{M_x\}$, $x\in\{1,...,k\}$, and an integer $n \geq 1$, we define the family of sets of subchannels given by $\Psi^{(\rho,\mathbb{M},n)}=\{\Psi^{(\rho,\mathbb{M},n)}_y(\cdot)\}$, $y\in\{1,...,k+n\}$ with:
\begin{align}
    \Psi^{(\rho,\mathbb{M},n)}_y(\eta) \coloneqq
    \bigg \{
    \begin{matrix}
    \alpha \Tr[Z^\rho \eta] Z^\mathbb{M}_y, 
    & \hspace{-1.1cm}y=1,...,k\\
    \frac{1}{n}
    [1-F(\eta)]\xi, 
    & \hspace{0.1cm} y=k+1,...,k+n
    \end{matrix}
    \label{eq:scgameD}
\end{align} 
with $\xi$ begin an arbitrary quantum state $\xi\geq 0$, $\Tr(\xi)=1$. We can check that this is a well-defined set of subchannels, because they now add up to a CPTP linear map:
\begin{align*}
    \Tr\left[
    \sum_{y=1}^{k+n} \Psi^{(\rho,\mathbb{M},n)}_y(\eta)
    \right]=1,
    \hspace{0.5cm}
    \forall n, \forall \eta.
\end{align*}
We now analyse the multi-object subchannel discrimination game given by $\Psi^{(\rho,\mathbb{M},n)}$. The probability of success of a player using the state-measurement pair $(\rho,\mathbb{M})$ is given by: 
\begin{align}
    P^{\rm D}_{\rm succ}(\Psi^{(\rho,\mathbb{M},n)}\rho,\mathbb{M})
    &= \nonumber
    \max_{\mathbb{N}\preceq \mathbb{M}}
    \sum_{y=1}^{k+n} 
    \Tr[N_y \Psi^{(\rho,\mathbb{M},n)}_y(\rho)]\\ \nonumber
    &\geq \sum_{x=1}^{k} 
    \Tr[M_x \Psi^{(\rho,\mathbb{M},n)}_x(\rho)] \\ \label{eq:startingpointD}
    & = \alpha \Tr[Z^\rho \rho] 
    \sum_{x=1}^{k} \Tr\left [
    M_x Z^\mathbb{M}_x
    \right].
\end{align}
The inequality follows because we have chosen to simulate a particular measurement, i.e. $N_y = M_y$ for $y \leq k$ and $N_y = 0$ for $y > k$. In the last equality we have replaced the subchannel discrimination game with \eqref{eq:scgameD}. Now, because of the conditions in \eqref{eq:condition1}, we have the \emph{strict} inequality:
\begin{align}
    P^{\rm D}_{\rm succ}(\Psi^{(\rho,\mathbb{M},n)},\rho,\mathbb{M})
    > 
    \alpha.
    \label{eq:ineqD1}
\end{align}
We now analyse the best \emph{fully free} player:
\begin{align*}
    \max_{
    \substack{
    \sigma \in {\rm F}\\
    \mathbb{N} \in \mathbb{F}
    }
    }
    P^{\rm D}_{\rm succ}(\Psi^{(\rho,\mathbb{M},n)},\sigma,\mathbb{N})
    =
    \max_{
    \substack{
    \sigma \in {\rm F}\\
    \mathbb{N} \in \mathbb{F}\\
    \mathbb{\tilde N} \preceq \mathbb{N}
    }
    }
    \sum_{x=1}^{k+n}\Tr
    \left[
    \tilde N_x \Psi^{(\rho,\mathbb{M},n)}_x(\sigma)
    \right].
\end{align*} 
We are considering QRTs of measurements for which CPP is a free operation and therefore, CPP is redundant here and we have:
\begin{align*}
    \max_{
    \substack{
    \sigma \in {\rm F}\\
    \mathbb{N} \in \mathbb{F}
    }
    }
    P^{\rm D}_{\rm succ}(\Psi^{(\rho,\mathbb{M},n)},\sigma,\mathbb{N})
    =
    \max_{
    \substack{
    \sigma \in {\rm F}\\
    \mathbb{N} \in \mathbb{F}
    }
    }
    \sum_{x=1}^{k+n}\Tr
    \left[
    N_x \Psi^{(\rho,\mathbb{M},n)}_x(\sigma)
    \right].
\end{align*} 
Let us now consider, without loss of generality, that these two maximisations are being achieved by the \emph{fully free} pair $(\sigma^*,\mathbb{N}^*)$. We then have:
\begin{multline}
    P^{\rm D}_{\rm succ} (\Psi^{(\rho,\mathbb{M},n)},\sigma^*,\mathbb{N}^*)
    =
    \sum_{x=1}^{k+n}\Tr
    \left[
    N^*_x \Psi^{(\rho,\mathbb{N},n)}_x(\sigma^*)
    \right],\\
    =
    \alpha \Tr[Z^\rho \sigma^*] 
    \sum_{y=1}^{k} \Tr
    \left[
    N^*_y Z^\mathbb{M}_y
    \right]\\
    +\frac{1}{n}\left[
    1-F(\sigma^*)
    \right]
    \sum_{y=k+1}^{k+n} \Tr\left[N^*_y\xi\right].
    \label{eq:eq10}
\end{multline}
In the second equality we have replaced the subchannel game \eqref{eq:scgameD}. The first term can be upper bounded as:
\begin{align*}
    \sum_{y=1}^{k} \Tr\left[
    N_y^* Z^\mathbb{M}_y
    \right
    ]\leq \sum_{y=1}^{k} \Tr
    \left[
    \tilde N^*_y Z^\mathbb{M}_y
    \right]
    \leq
    1,
\end{align*}
with the measurement $\mathbb{\tilde N}^*$ defined in \eqref{eq:Ntilde}. The first inequality follows from the definition of the measurement $\mathbb{\tilde N}^*$.  In the second inequality we use the fact that $\mathbb{\tilde N}^*$ is a free measurement (because it was constructed from a free measurement $\mathbb{N}^*$ and a CPP operation, which is a free operation) and therefore we can use the conditions in \eqref{eq:condition2}. We now also use the fact that $1-F(\eta)\leq 1$, $\forall \eta$, then equation \eqref{eq:eq10} becomes:
\begin{align*}
    P^{\rm D}_{\rm succ}(\Psi^{(\rho,\mathbb{M},n)},\sigma,\mathbb{N})
    &\leq \alpha + 
    \frac{1}{n}
    \sum_{y=k+1}^{k+n} \Tr\left[N^*_y\xi\right].
\end{align*}
The second term can be upper bounded as:
\begin{align*}
    \sum_{y=k+1}^{k+n} \Tr[N^*_y\xi] 
    \leq  
    \sum_{y=1}^{k+n} \Tr[N^*_y\xi]&=
    \Tr \left[\left (\sum_{y=1}^{k+n} N^*_y \right)\xi\right]=1.
\end{align*}
The inequality follows because we have added positive terms and the equality follows from $\mathbb{N}^*$ being a measurement $\sum_{y=1}^{k+n} \tilde N_y=\mathds{1}$ and $\xi$ being a state. We then get:
\begin{align*}
    &P^{\rm D}_{\rm succ}(\Psi^{(\rho,\mathbb{M},n)},\sigma,\mathbb{N})
    \leq 
    \alpha +\frac{1}{n}.
\end{align*}
We now choose the subchannel game given by $\Psi^{(\rho,\mathbb{M},n\rightarrow \infty)}$ and therefore we get:
\begin{align}
    &P^{\rm D}_{\rm succ}
    \left(
    \Psi^{(\rho,\mathbb{M},n\rightarrow \infty)},\sigma,\mathbb{N}
    \right)
    \leq 
    \alpha.
    \label{eq:ineqD2}
\end{align}
Finally, equations \eqref{eq:ineqD1} and \eqref{eq:ineqD2} together imply that:
\begin{align*}
    \max_{
    \substack{
    \sigma \in {\rm F}\\
    \mathbb{N} \in \mathbb{F}
    }
    }
    P^{\rm D}_{\rm succ}
    (
    \Psi^{(\rho, \mathbb{M}, n\rightarrow \infty)},\sigma,\mathbb{N}
    )
    <
    P^{\rm D}_{\rm succ}
    (
    \Psi^{(\rho, \mathbb{M}, n\rightarrow \infty)},\rho,\mathbb{M}
    ),
\end{align*}
as desired.
\end{proof}

\subsection{Exclusion case}

\textbf{Result 1B:} Consider a convex QRT of states with an arbitrary resource and a convex QRT of measurements with an arbitrary resource for which CPP is a free operation. Given any \emph{fully resourceful} state-measurement pair, meaning that we have a resourceful state $\rho \notin {\rm F}$ and a resourceful measurement $\mathbb{M}\notin\mathbb{F}$, then, there exist a subchannel game $\Psi^{(\rho,\mathbb{M})}$ such that:
\begin{align}
    P^{\rm E}_{\rm err}(\Psi^{(\rho,\mathbb{M})},\rho,\mathbb{M})
    <
    \min_{\sigma \in {\rm F}}
    \min_{\mathbb{N} \in \mathbb{F}}
    P^{\rm E}_{\rm err}(\Psi^{(\rho,\mathbb{M})},\sigma,\mathbb{N}),
    \label{eq:result1EA}
\end{align}
with minimisation over all possible free states and measurements.

\begin{proof}
This proof is closely related to the discrimination proof, but the subchannel game has to be constructed differently. We start by considering a \emph{fully resourceful} state-measurement pair $(\rho,\mathbb{M})$. Using the hyperplane separation theorem \cite{CA_book}, there exist positive semidefinite operators $Y^\rho$ and $\{Y^\mathbb{M}_x\}$, $x\in\{1,...,k\}$ satisfying the conditions \eqref{eq:condition3} and \eqref{eq:condition4}. We now define the set of maps $\{\Phi^{(\rho,\mathbb{M})}_x (\cdot)\}$ with:
\begin{align}
    \Phi^{(\rho,\mathbb{M})}_x(\eta)
    &\coloneqq
    \beta^{(\rho,\mathbb{M})}
    \Tr(Y^\rho \eta)
    Y^\mathbb{M}_x, \nonumber\\
    \beta^{(\rho,\mathbb{M})}
    &\coloneqq 
    \frac{1}{
    2
    \norm{Y^\rho}_1
    \Tr(Y^\mathbb{M})
    },
    \hspace{0.5cm}
    Y^\mathbb{M} \coloneqq
    \sum_{x=1}^{k} Y^\mathbb{M}_x.
    \label{eq:beta}
\end{align}
with $\norm{X}_1=\Tr(\sqrt{X^\dagger X})$ the trace norm. We are going to use the notation $\beta=\beta^{(\rho,\mathbb{M})}$. As before, these operators are completely-positive linear maps and they now satisfy that $\forall \eta$:
\begin{align*}
    G(\eta) 
    \coloneqq 
    \Tr \left[
    \sum_{x=1}^{k} \Phi^{(\rho,\mathbb{M})}_x(\eta)
    \right]=
    \frac{\Tr(Y^\rho \eta)}{2\norm{Y^\rho}_1}
    \leq 
    \frac{1}{2},
\end{align*}
which can also be written as:
\begin{align}
    G(\eta)=\beta\Tr(Y^\rho \eta)\Tr (Y^\mathbb{M}).
    \label{eq:G}
\end{align}
The set of maps $\{\Phi^{(\rho,\mathbb{M})}_x(\cdot)\}$ then add up to a completely positive trace-nonincreasing linear map. We can then \emph{complete} this set to be a set of \emph{subchannels} by adding an extra subchannel $\Psi^{(\rho, \mathbb{M})}_{k+1}(\cdot) \coloneqq \Lambda(\cdot)- \Phi^{(\rho,\mathbb{M})}(\cdot)$, with $\Lambda$ being an arbitrary CPTP map such that it is greater or equal than zero (take the identity channel for instance). Therefore, with this construction we obtain a well-defined set of subchannels with $k+1$ elements. We now proceed to define a set of subchannels in the following manner. Given a state-measurement pair $(\rho,\mathbb{M})$, $\mathbb{M}=\{M_x\}$, $x\in\{1,...,k\}$, we define the set of subchannels given by $\Psi^{(\rho,\mathbb{M})}=\{\Psi^{(\rho,\mathbb{M})}_y(\cdot)\}$, $y\in\{1,...,k+n\}$ with:
\begin{align}
    \Psi^{(\rho,\mathbb{M})}_y(\eta) 
    \coloneqq
    \bigg \{
    \begin{matrix}
    \beta \Tr[Y^\rho \eta] Y^\mathbb{M}_y, 
    & \hspace{0.3cm}y=1,...,k\\
    [1-G(\eta)]\xi^\mathbb{M}, 
    & \hspace{0.2cm} y=k+1
    \end{matrix}
    \label{eq:scgameE}
\end{align} 
with the quantum state: 
\begin{align}
    \xi^\mathbb{M}
    \coloneqq
    \frac{
    \sum_{x=1}^{k} p(x) Y^\mathbb{M}_x.
    }{
    \sum_{x=1}^{k} p(x) \Tr \left(Y^\mathbb{M}_x \right)
    }.
    \label{eq:xiM}
\end{align}
$\{p(x)\}$ being an arbitrary probability distribution. We can also check that this is a well-defined set of subchannels, i. e., they add up to a CPTP linear map:
\begin{align*}
    \Tr\left[
    \sum_{y=1}^{k+1} \Psi^{(\rho,\mathbb{M})}_y(\eta)
    \right]=1,
    \hspace{0.5cm}
    \forall \eta.
\end{align*}
We remark here that, unlike the discrimination case, we are not generating a family of sets of subchannels, but only a specific one. We now analyse the multi-object subchannel exclusion game given by $\Psi^{(\rho,\mathbb{M})}$ and the probability of error of a player using the state-measurement pair $(\rho,\mathbb{M})$ which is given by: 
\begin{align}
    P^{\rm E}_{\rm err}(\Psi^{(\rho,\mathbb{M})}\rho,\mathbb{M})
    &= \nonumber
    \min_{\mathbb{N}\preceq \mathbb{M}}
    \sum_{y=1}^{k+n} 
    \Tr[N_y \Psi^{(\rho,\mathbb{M})}_y(\rho)]\\ \nonumber
    &\leq \sum_{x=1}^{k} 
    \Tr[M_x \Psi^{(\rho,\mathbb{M})}_x(\rho)] \\ \label{eq:startingpointE}
    & = \beta \Tr[Y^\rho \rho] 
    \sum_{x=1}^{k} \Tr\left [
    M_x Y^\mathbb{M}_x
    \right].
\end{align}
The inequality follows because we have chosen to simulate a particular measurement, i.e. $N_y = M_y$ for $y \leq k$ and $N_y = 0$ for $y > k$. In the last equality we have replaced the subchannel exclusion game with \eqref{eq:scgameE}. Now, because of \eqref{eq:condition3} and \eqref{eq:condition4}, we have the \emph{strict} inequality:
\begin{align}
    P^{\rm E}_{\rm err}(\Psi^{(\rho,\mathbb{M})},\rho,\mathbb{M})
    < 
    \beta.
    \label{eq:ineqE1}
\end{align}
As before, we now analyse the best \emph{fully free} player:
\begin{align*}
    \min_{
    \substack{
    \sigma \in {\rm F}\\
    \mathbb{N} \in \mathbb{F}
    }
    }
    P^{\rm E}_{\rm err}(\Psi^{(\rho,\mathbb{M})},\sigma,\mathbb{N})
    &=
    \min_{
    \substack{
    \sigma \in {\rm F}\\
    \mathbb{N} \in \mathbb{F}\\
    \mathbb{\tilde N} \preceq \mathbb{N}
    }
    }
    \sum_{x=1}^{k+1}\Tr
    \left[
    \tilde N_x \Psi^{(\rho,\mathbb{M})}_x(\sigma)
    \right]\\
    &=
    \min_{
    \substack{
    \sigma \in {\rm F}\\
    \mathbb{N} \in \mathbb{F}
    }
    }
    \sum_{x=1}^{k+1}\Tr
    \left[
    N_x \Psi^{(\rho,\mathbb{M})}_x(\sigma)
    \right],
\end{align*} 
where the equality follows because CPP is redundant. Let us now consider, without loss of generality, that these two minimisations are achieved by the \emph{fully free} pair $(\sigma^*,\mathbb{N}^*)$. We then have:
\begin{align*}
    P^{\rm E}_{\rm err}(\Psi^{(\rho,\mathbb{M})},\sigma^*,\mathbb{N}^*)
    =&\sum_{x=1}^{k+1} 
    \Tr[N^*_x \Psi^{(\rho,\mathbb{N})}_x(\sigma^*)]\\
    =&
    \beta \Tr[Y^\rho \sigma] 
    \sum_{y=1}^{k} \Tr\left [
    N^*_y Y^\mathbb{M}_y
    \right]\\
    &+ \left[
    1-G(\sigma^*)
    \right]
    \Tr[N^*_{k+1}\xi^\mathbb{M}].
\end{align*}
We now add and subtract a convenient term as:
\begin{align*}
    P^{\rm E}_{\rm err}(\Psi^{(\rho,\mathbb{M})},\sigma^*,\mathbb{N}^*)
    &=
    \beta \Tr[Y^\rho \sigma] 
    \sum_{x=1}^{k} \Tr\left [
    N^*_x Y^\mathbb{M}_x
    \right]\\
    &+\beta \Tr (Y^\rho \sigma^*)
    \sum_{x=1}^k p(x)\Tr (N^*_{k+1}Y^\mathbb{M}_x)\\
    &-\beta \Tr (Y^\rho \sigma^*)
    \sum_{x=1}^k p(x)\Tr (N^*_{k+1}Y^\mathbb{M}_x)\\
    &+ 
    \left[
    1-G(\sigma^*)
    \right]
    \Tr \left[N^*_{k+1}\xi^\mathbb{M}\right].
\end{align*}
We now define a measurement given by $\mathbb{\tilde N}=\{\tilde N^*_x\}$ with $\tilde N^*_x=N^*_x+p(x)N^*_{k+1}$, and $p(x)$ being the probability distribution from \eqref{eq:xiM}, and we can reorganise this as:
\begin{align*}
    P^{\rm E}_{\rm err}(\Psi^{(\rho,\mathbb{M})},\sigma^*,\mathbb{N}^*)
    &=
    \beta \Tr[Y^\rho \sigma] 
    \sum_{y=1}^{k} \Tr\left [
    \tilde N^*_y Y^\mathbb{M}_y
    \right]\\
    &+
    \left[
    1-G(\sigma^*)
    \right]
    \Tr \left[N^*_{k+1}\xi^\mathbb{M}\right]\\
    &-\beta \Tr (Y^\rho \sigma^*)
    \sum_{x=1}^k p(x)\Tr (N^*_{k+1}Y^\mathbb{M}_x).
\end{align*} 
The first term is lower bounded by $\beta$ by using the conditions in \eqref{eq:condition4} and therefore we have:
\begin{multline}
    P^{\rm E}_{\rm err}
    (\Psi^{(\rho,\mathbb{M},n)},\sigma^*,\mathbb{N}^*) \geq
    \beta\\
    +
    \left[
    1-G(\sigma^*)
    \right]
    \Tr \left[N^*_{k+1}\xi^\mathbb{M}\right]\\
    -\beta \Tr (Y^\rho \sigma^*)
    \sum_{x=1}^k p(x) \Tr (N^*_{k+1}Y^\mathbb{M}_x).
    \label{eq:ineqE2}
\end{multline}
We now prove that the remaining term (last two lines) is always greater than or equal to zero. We start by rewriting this term as:
\begin{multline}
    \nonumber
    \left[
    1-G(\sigma^*)
    \right]
    \Tr \left[N^*_{k+1}\xi^\mathbb{M}\right]\\
    -
    \beta \Tr (Y^\rho \sigma^*)
    \sum_{x=1}^k p(x)
    \Tr (N^*_{k+1}Y^\mathbb{M}_x)\\
    =
    \Tr
    \Bigg\{
    N^*_{k+1}
    \Big[
    \left(
    1-G(\sigma^*)
    \right)
    \xi^\mathbb{M}
    -\beta \Tr (Y^\rho \sigma^*)
    \sum_{x=1}^k p(x)Y^\mathbb{M}_x
    \Big]
    \Bigg\}.
\end{multline}
We have $N^*_{k+1}\geq 0$ and therefore we now only need to prove that the operator inside the square brackets is positive semidefinite. We rewrite this operator as:
\begin{align*}
    &\left[
    1-G(\sigma^*)
    \right]
    \xi^\mathbb{M}
    -\beta \Tr (Y^\rho \sigma^*)
    \sum_{x=1}^k p(x)Y^\mathbb{M}_x\\
    =&\left[
    1-G(\sigma^*)
    \right]
    \frac{
    \sum_{x=1}^k p(x)Y^\mathbb{M}_x
    }{
    \sum_{x=1}^k p(x)\Tr (Y^\mathbb{M}_x)
    }
    -
    \beta \Tr (Y^\rho \sigma^*)
    \sum_{x=1}^k p(x)Y^\mathbb{M}_x,
\end{align*}
where we used \eqref{eq:xiM} to substitute for $\xi^\mathbb{M}$. We now multiply by the positive term $\sum_{x=1}^k p(x)\Tr (Y^\mathbb{M}_x)$ and obtain:
\begin{align*}
    &\left[
    1-G(\sigma^*)
    \right]
    \sum_{x=1}^k p(x)Y^\mathbb{M}_x\\
    &-
    \beta \Tr (Y^\rho \sigma^*)
    \left(
    \sum_{x=1}^k p(x)\Tr (Y^\mathbb{M}_x)
    \right)
    \left(
    \sum_{x=1}^k p(x)Y^\mathbb{M}_x
    \right).
\end{align*}
We now factorise the positive semidefinite operator $\sum_{x=1}^k p(x)Y^\mathbb{M}_x$ and analyse the coefficient as follows:
\begin{multline}
    1-G(\sigma^*)
    -
    \beta \Tr (Y^\rho \sigma^*)
    \sum_{x=1}^k p(x)\Tr (Y^\mathbb{M}_x)\\
    =1-
    \beta 
    \Tr (Y^\rho \sigma^*)
    \Tr (Y^\mathbb{M})
    -
    \beta 
    \Tr (Y^\rho \sigma^*)
    \sum_{x=1}^k p(x)\Tr (Y^\mathbb{M}_x),\\
    \geq
    1
    -
    2\beta 
    \Tr (Y^\rho \sigma^*)
    \Tr (Y^\mathbb{M})
    =
    1
    -
    \frac{ 
    \Tr (Y^\rho \sigma^*)
    }
    {\norm{Y^\rho}_1
    }
    \geq
     0.
\end{multline}
In the first equality we replaced $G(\sigma^*)$ using \eqref{eq:G}. The first inequality follows because we are subtracting a larger quantity. In the second equality we substituted $\beta$ \eqref{eq:beta}. The second inequality follows because $\frac{\Tr (Y^\rho \eta)}{\norm{Y^\rho}_1}\leq 1$, $\forall \eta$. Coming back to \eqref{eq:ineqE2} we then have:
\begin{align}
    &P^{\rm E}_{\rm err}
    (\Psi^{(\rho,\mathbb{M})},\sigma^*,\mathbb{N}^*) \geq
    \beta.
    \label{eq:ineqE3}
\end{align}
Putting together \eqref{eq:ineqE1} and \eqref{eq:ineqE3} we obtain:
\begin{align*}
    P^{\rm E}_{\rm err}
    (
    \Psi^{(\rho, \mathbb{M})},\rho,\mathbb{M}
    )
    <
    \min_{\sigma \in {\rm F}}
    \min_{\mathbb{N} \in \mathbb{F}}
    P^{\rm E}_{\rm err}
    (
    \Psi^{(\rho, \mathbb{M})},\sigma,\mathbb{N}
    ),
\end{align*}
as desired.
\end{proof}

\newpage
\section{Proof of Result 2}

We divide this result in two parts. In the first part we prove the upper bound for discrimination and the lower bound for exclusion. In the second part, we show how to achieve these bounds.

\textbf{Result 2A:} Consider a convex QRT of states with an arbitrary resource and a convex QRT of measurements with an arbitrary resource for which CPP is a free operation. Given \emph{any} state-measurement pair $(\rho, \mathbb{M})$ we have:
\begin{align}
    \hspace{-0.2cm}\max_{\rm \Psi} \frac{
    P^{\rm D}_{\rm succ}(\Psi,\rho,\mathbb{M})
    }{
    \displaystyle
    \max_{
    \substack{
    \sigma \in {\rm F}\\
    \mathbb{N} \in \mathbb{F}
    }
    }
    P^{\rm D}_{\rm succ}(\Psi,\sigma,\mathbb{N})
    } \hspace{-0.1cm}
    = \hspace{-0.1cm}
    \Big [1+{\rm R_F}(\rho)\Big]
    \Big[1+{\rm R_\mathbb{F}}(\mathbb{M}) \Big].
    \label{eq:result2RA}
\end{align}

\textbf{Result 2B:} Consider a convex QRT of states with an arbitrary resource and a convex QRT of measurements with an arbitrary resource for which CPP is a free operation. Given \emph{any} state-measurement pair $(\rho, \mathbb{M})$ we have:
\begin{align}
    \hspace{-0.15cm}\min_{\rm \Psi} \frac{
    P^{\rm E}_{\rm err}(\Psi,\rho,\mathbb{M})
    }{
    \displaystyle
    \min_{
    \substack{
    \sigma \in {\rm F}\\
    \mathbb{N} \in \mathbb{F}
    }
    }
    P^{\rm E}_{\rm err}(\Psi,\sigma,\mathbb{N})
    }\hspace{-0.1cm} = 
    \hspace{-0.1cm}
    \Big [1-{\rm W_F}(\rho)\Big]
    \Big[1-{\rm W_\mathbb{F}}(\mathbb{M}) \Big].
    \label{eq:result2WA}
\end{align}

\begin{widetext}
\subsection{Upper bound for multi-object discrimination and lower bound for multi-object exclusion}

We start by proving that for \emph{any} state-measurement pair $(\rho,\mathbb{M})$, the product $[1+{\rm R_F}(\rho)][1+{\rm R_\mathbb{F}}(\mathbb{M})]$ places an upper bound on the advantage ratio in any subchannel game $\Psi$. 

\begin{proof}
Given \emph{any} subchannel game $\Psi$ and \emph{any} state-measurement pair $(\rho,\mathbb{M})$ we have:
\begin{align}
    P^{\rm D}_{\rm succ}(\Psi,\rho,\mathbb{M})
    =
    \max_{\mathbb{N}\preceq \mathbb{M}}
    \sum_{x}\Tr[N_x\Psi_x(\rho)] 
    &\leq 
    \Big[1+{\rm R_F}(\rho)\Big] 
    \max_{\mathbb{N}\preceq \mathbb{M}}
    \sum_{x} {\rm Tr}[N_x\Psi_x(\sigma^*)], \nonumber\\
    &\leq \Big[1+{\rm R_F}(\rho)\Big]
    \max_{\sigma \in {\rm F}}
    \max_{\mathbb{N}\preceq \mathbb{M}}
    \sum_{x} {\rm Tr}[N_x\Psi_x(\sigma)], \nonumber\\
    & = \Big[1+{\rm R_F}(\rho)\Big]
    \max_{\sigma \in {\rm F}}
    \max_{\{q(x|a)\}}
    \sum_{x} {\rm Tr}
    \left[
    \left(
    \sum_a q(x|a)M_a
    \right)
    \Psi_x(\sigma)
    \right], \nonumber\\
    &\leq 
    \Big[1+{\rm R_F}(\rho)\Big]
    \Big[1+{\rm R_\mathbb{F}}(\mathbb{M})\Big] 
    \max_{\sigma \in {\rm F}}
    \max_{\{q(x|a)\}}
    \sum_{x} {\rm Tr}
    \left[
    \left(
    \sum_a q(x|a)\tilde N^*_a
    \right)
    \Psi_x(\sigma)
    \right], \nonumber\\
    & =
    \Big[1+{\rm R_F}(\rho)\Big]
    \Big[1+{\rm R_\mathbb{F}}(\mathbb{M})\Big]
    \max_{\sigma \in {\rm F}}
    \max_{\mathbb{\dbtilde{N}}\preceq \mathbb{\tilde N}^*}
    \sum_{x} {\rm Tr}
    \left[\dbtilde{N}_x\Psi_x(\sigma)
    \right], \nonumber\\
    &\leq 
    \Big[1+{\rm R_F}(\rho)\Big]
    \Big[1+{\rm R_\mathbb{F}}(\mathbb{M})\Big]
    \max_{\sigma \in {\rm F}}
    \max_{\mathbb{\tilde N}\in \mathbb{F}}
    \max_{\mathbb{\dbtilde{N}}\preceq \mathbb{\tilde N}}
    \sum_{x} {\rm Tr}
    \left[
    \dbtilde{N}_x\Psi_x(\sigma)
    \right], \nonumber\\
    &=
    \Big[1+{\rm R_F}(\rho)\Big]
    \Big[1+{\rm R_\mathbb{F}}(\mathbb{M})\Big]
    \max_{\sigma \in {\rm F}}
    \max_{\mathbb{\tilde N}\in \mathbb{F}}
    P^{\rm D}_{\rm succ}(\Psi,\sigma,\mathbb{\tilde{N}}).
    \label{eq:upperbound}
\end{align}
In the first inequality we use the definition of the generalised robustness from which we get $\rho \leq [1+{\rm R_F}(\rho)]\sigma^*$ and since $\Psi_x$ are linear maps we have $\Psi_x(\rho)\leq [1+{\rm R_F}(\rho)] \Psi_x(\sigma^*)$, $\forall x$. In the second inequality we allow ourselves to maximise over all free states. In the third inequality, we use the definition of the generalised robustness from which we get $M_a \leq [1+{\rm R_\mathbb{M}}(\mathbb{M})]\tilde N_a^*$, $\forall a$. In the fourth inequality we allow ourselves to maximise over all free measurements.
\end{proof}
\end{widetext}

\begin{proof}
The proof for the lower bound for multi-object subchannel exclusion follows similar arguments.
\end{proof}

\subsection{Achieving upper bound for discrimination and lower bound for exclusion}

\textbf{Lemma 1:} (Dual SDPs for generalised robustness) The generalised robustness of resource of a state $\rho$ and a measurement $\mathbb{M}=\{M_x\}$, $x\in\{1,...,k\}$ can be written as:
\begin{subequations}
\begin{align}
    {\rm R_F}\left(\rho\right)=
    \max_{Z} \hspace{0.2cm}
    &{\rm Tr}[(Z)\rho]-1, \label{eq:RoRs1}\\
    {\rm s. t.} \hspace{0.2cm}
    & Z\geq 0,\label{eq:RoRs2}\\
    &{\rm Tr}[Z\sigma]\leq 1, \hspace{0.4cm} \forall \sigma \in {\rm F} ,
    \label{eq:RoRs3}
\end{align}
\end{subequations}
\vspace{-0.7cm}
\begin{subequations}
\begin{align}
    {\rm R_\mathbb{F}}\left(\mathbb{M}\right)=
    \max_{\{Z_x\}} \hspace{0.2cm}
    &\sum_x{\rm Tr}[Z_xM_x]-1, \label{eq:RoRm1}\\
    {\rm s. t.} \hspace{0.2cm}
    & Z_x\geq 0, \hspace{0.5cm}\forall x, \label{eq:RoRm2}\\
    \sum_x\Tr[& Z_x N_x]\leq 1, \hspace{0.2cm} \forall \mathbb{N} \in \mathbb{F}.
    \label{eq:RoRm3}
\end{align}
\end{subequations}
These are the dual SDP formulations of the generalised robustnesses of resource for states and measurements.

\begin{proof}\emph{(of Result 2A)} Given \emph{any} state-measurement pair $(\rho,\mathbb{M})$, we want to find a suitable subchannel game $\Psi$ so that we achieve the upper bound in \eqref{eq:upperbound}. We start by noting that Lemma 1 is a refined version of the hyperplane separation theorem, from which we can extract positive semidefinite operators $Z^\rho$, $\{Z^\mathbb{M}_x\}$, $x\in\{1,...,k\}$ satisfying the conditions \eqref{eq:condition1} and \eqref{eq:condition2}. Therefore, the construction of the set of subchannels from the previous section applies here as well. We then continue from \eqref{eq:startingpointD} which can now be rewritten as:
\begin{multline}
    P^{\rm D}_{\rm succ}(\Psi^{(\rho,\mathbb{M},n)},\rho,\mathbb{M})
    \geq
    \alpha \Tr[Z^\rho \rho] 
    \sum_{y=1}^{k} \Tr\left [
    M_y Z^\mathbb{M}_y
    \right],\\
    =\alpha 
    \Big[ 1+{\rm R_F}(\rho) \Big]
    \Big[ 1+{\rm R_\mathbb{F}}(\mathbb{M}) \Big].
    \label{eq:b4r0}
\end{multline}
The equality follows from \eqref{eq:RoRs1} and \eqref{eq:RoRm1}. We now analyse the \emph{fully free} player. Similarly, we now choose the subchannel game given by $\Psi^{(\rho,\mathbb{M},n\rightarrow \infty)}$ and invoking \eqref{eq:ineqD2} we have:
\begin{align}
    \max_{\sigma \in {\rm F}}
    \max_{\mathbb{N} \in \mathbb{F}}
    P^{\rm D}_{\rm succ}
    (
    \Psi^{(\rho,\mathbb{M},n\rightarrow \infty)},\sigma,\mathbb{N}
    )
    \leq 
    \alpha.
    \label{eq:b4r1}
\end{align}
We now analyse the ratio of interest with this particular subchannel game and have:
\begin{multline}
    \frac{
    P^{\rm D}_{\rm succ}
    \left(
    \Psi^{(\rho,\mathbb{M},n\rightarrow \infty)},\rho,\mathbb{M}
    \right)
    }{
    \displaystyle
    \max_{\sigma \in {\rm F}}
    \max_{\mathbb{N} \in \mathbb{F}}
    P^{\rm D}_{\rm succ}
    \left(
    \Psi^{(\rho,\mathbb{M},n\rightarrow \infty)},\sigma,\mathbb{N}
    \right)
    }\\
    \geq 
    \frac{
    \alpha \Big[ 1+{\rm R_F}(\rho) \Big]
    \Big[ 1+{\rm R_\mathbb{F}}(\mathbb{M}) \Big]
    }{
    \displaystyle
    \max_{\sigma \in {\rm F}}
    \max_{\mathbb{N} \in \mathbb{F}}
    P^{\rm D}_{\rm succ}
    \left(\Psi^{(\rho,\mathbb{M},n\rightarrow \infty)},\sigma,\mathbb{N}
    \right)
    }\\
    \geq \frac{
    \alpha 
    \Big[ 1+{\rm R_F}(\rho) \Big]
    \Big[ 1+{\rm R_\mathbb{F}}(\mathbb{M}) \Big]
    }{
    \alpha
    }\\
    =\Big[ 1+{\rm R_F}(\rho) \Big]
    \Big[ 1+{\rm R_\mathbb{F}}(\mathbb{M}) \Big].
    \label{eq:b4r2}
\end{multline}
In the first inequality we used \eqref{eq:b4r0} whilst in the second we used \eqref{eq:b4r1}. Putting together \eqref{eq:b4r2} and \eqref{eq:upperbound} we obtain:
\begin{align*}
    \frac{
    P^{\rm D}_{\rm succ}(\Psi^{(\rho,\mathbb{M},n\rightarrow \infty)},\rho,\mathbb{M})
    }{
    \displaystyle
    \max_{
    \substack{
    \sigma \in {\rm F}\\
    \mathbb{N} \in \mathbb{F}
    }
    }
    P^{\rm D}_{\rm succ}(\Psi^{(\rho,\mathbb{M},n\rightarrow \infty)},\sigma,\mathbb{N})
    }
    = \hspace{-0.1cm}
    \Big[ 1+{\rm R_F}(\rho) \Big]
    \Big[ 1+{\rm R_\mathbb{F}}(\mathbb{M}) \Big].
\end{align*}
as desired.
\end{proof}

\textbf{Lemma 2:} (Dual SDPs for weight) The weight of resource of a state $\rho$ and a measurement $\mathbb{M}=\{M_x\}$, $x\in\{1,...,k\}$ can be written as:
\begin{subequations}
\begin{align}
    {\rm W_F}\left(\rho\right)=
    \max_{Y} \hspace{0.2cm}
    &{\rm Tr}[(-Y)\rho]+1, \label{eq:WoRs1}\\
    {\rm s. t.} \hspace{0.2cm}
    & Y\geq 0,\label{eq:WoRs2}\\
    &{\rm Tr}[Y\sigma]\geq 1, \hspace{0.4cm} \forall \sigma \in {\rm F},
    \label{eq:WoRs3}
\end{align}
\end{subequations}
\vspace{-0.7cm}
\begin{subequations}
\begin{align}
    {\rm W_\mathbb{F}}\left(\mathbb{M}\right)=
    \max_{\{Y_x\}} \hspace{0.2cm}
    &\sum_x{\rm Tr}[(-Y_x)M_x]+1, \label{eq:WoRm1}\\
    {\rm s. t.} \hspace{0.2cm}
    & Y_x\geq 0, \hspace{0.5cm}\forall x, \label{eq:WoRm2}\\
    \sum_x \Tr[& Y_x N_x] \geq 1, \hspace{0.2cm} \forall \mathbb{N} \in \mathbb{F}.
    \label{eq:WoRm3}
\end{align}
\end{subequations}
These are the dual SDP formulations of the weights of resource for states and measurements.

\begin{proof}\emph{(of Result 2B)} This proof follows a similar logic to that of the robustness, and we write down for completeness. Given \emph{any} state-measurement pair $(\rho,\mathbb{M})$, we want to find a suitable subchannel game $\Psi$ so that we achieve the lower bound in \eqref{eq:result2WA}. The construction of the set of subchannels form the previous section applies here as well. We then continue from \eqref{eq:startingpointE} which can now be rewritten as:
\begin{multline}
    P^{\rm E}_{\rm err}(\Psi^{(\rho,\mathbb{M})}\rho,\mathbb{M})
    \leq
    \beta \Tr[Y^\rho \rho] 
    \sum_{y=1}^{k} \Tr\left [
    M_y Y^\mathbb{M}_y
    \right],\\
    =
    \beta
    \Big[ 1-{\rm W_F}(\rho) \Big]
    \Big[ 1-{\rm W_\mathbb{F}}(\mathbb{M}) \Big].
    \label{eq:b4w0}
\end{multline}
The equality follows from \eqref{eq:WoRs1} and \eqref{eq:WoRm1}. We now analyse the \emph{fully free} player and invoke \eqref{eq:ineqE3} which reads:
\begin{align}
    \min_{\sigma \in {\rm F}}
    \min_{\mathbb{N} \in \mathbb{F}}
    P^{\rm E}_{\rm err} (\Psi^{(\rho,\mathbb{M})},\sigma,\mathbb{N})
    \geq 
    \beta.
    \label{eq:b4w1}
\end{align}
We now analyse the ratio of interest with this particular subchannel game and have:
\begin{multline}
    \frac{
    P^{\rm E}_{\rm err}
    \left(
    \Psi^{(\rho,\mathbb{M})},\rho,\mathbb{M}
    \right)
    }{
    \displaystyle
    \min_{\sigma \in {\rm F}}
    \min_{\mathbb{N} \in \mathbb{F}}
    P^{\rm E}_{\rm err}
    \left(
    \Psi^{(\rho,\mathbb{M})},\sigma,\mathbb{N} 
    \right)
    }\\
    \leq 
    \frac{
    \beta 
    \Big[ 1-{\rm W_F}(\rho) \Big]
    \Big[ 1-{\rm W_\mathbb{F}}(\mathbb{M}) \Big]
    }{
    \displaystyle
    \min_{\sigma \in {\rm F}}
    \min_{\mathbb{N} \in \mathbb{F}}
    P^{\rm E}_{\rm err}
    \left(\Psi^{(\rho,\mathbb{M})},\sigma,\mathbb{N}
    \right)
    }\\
    \leq
    \frac{
    \beta 
    \Big[ 1-{\rm W_F}(\rho) \Big]
    \Big[ 1-{\rm W_\mathbb{F}}(\mathbb{M}) \Big]
    }{
    \beta
    }\\
    =\Big[ 1-{\rm W_F}(\rho) \Big]
    \Big[ 1-{\rm W_\mathbb{F}}(\mathbb{M}) \Big].
    \label{eq:b4w2}
\end{multline}
In the first inequality we used \eqref{eq:b4w0} whilst in the second we used \eqref{eq:b4w1}. Putting together \eqref{eq:b4w2} and the lower bound in \autoref{eq:result2WA} we obtain:
\begin{align*}
    \frac{
    P^{\rm E}_{\rm err} (\Psi^{(\rho,\mathbb{M})},\rho,\mathbb{M})
    }{
    \displaystyle
    \min_{
    \substack{
    \sigma \in {\rm F}\\
    \mathbb{N} \in \mathbb{F}
    }
    }
    P^{\rm E}_{\rm err} (\Psi^{(\rho,\mathbb{M})},\sigma,\mathbb{N})
    }
    =\Big[ 1-{\rm W_F}(\rho) \Big]
    \Big[ 1-{\rm W_\mathbb{F}}(\mathbb{M}) \Big].
\end{align*}
as desired.
\end{proof}

\section{Proof of Result 3}

\textbf{Result 3A:} The maximum gap between the order plus-infinity mutual information of \emph{any} state-measurement pair $(\rho,\mathbb{M})$ when compared to the best \emph{fully free} state-measurement pair is upper bounded as:
\begin{align}
    \nonumber 
    &\max_{\Lambda} 
    \left\{
    I_{+\infty}(X_{\Lambda,\rho} \colon G_{\mathbb{M}})
    -
    \max_{\sigma\in {\rm F}}
    \max_{\mathbb{N}\in \mathbb{F}}
    I_{+\infty}(X_{\Lambda,\sigma} \colon G_{\mathbb{N}})
    \right\}\\
    &\leq
    \log \Big[1+{\rm R_F}(\rho)\Big]
    +
    \log \Big[1+{\rm R_\mathbb{F}}(\mathbb{M})\Big],
    \label{eq:result3RA}
\end{align}
with the maximisation over all ensembles of channels. 

\begin{proof}
The plus-infinity mutual information between classical random variables $X_{\Lambda,\rho}$ and $G_\mathbb{M}$ is given by \cite{RR_thesis}:
\begin{align*}
    I_{+\infty}(X_{\Lambda,\rho} \colon G_\mathbb{M})=+
    \left[
    H_{+\infty}(X_{\Lambda,\rho})-
    H_{+\infty}(X_{\Lambda,\rho} |G_\mathbb{M})
    \right],
\end{align*}
with $H_{+\infty}(X_{\Lambda,\rho})=- \log (\max_x p(x))$, $H_{+\infty}(X_{\Lambda,\rho}|G_\mathbb{M})=- \log (\sum_g \max_x p(g,x))$ with $p(g,x)=p(g|x)p(x)$. We have $p(g|x)=\Tr(M_g \Lambda_x(\rho))$ and $H_{+\infty}(X_{\Lambda,\rho}|G_\mathbb{M})=- \log \sum_g \max_x \Tr[M_g\Lambda_x(\rho)]p(x)$. Considering $f_g(x)={\rm Tr}[M_g \Lambda_x(\rho)]p(x)$ and using:
\begin{align}
    \max_x f_g(x) =\max_{\{p(x|g)\}} \sum_x p(x|g) f_g(x),
    \label{eq:p1}
\end{align}
we have:
\begin{align} 
    \nonumber & H_{+\infty}(X_{\Lambda,\rho}|G_\mathbb{M})\\
    &=- \log 
    \nonumber \sum_g \max_{\{p(x|g)\}} \sum_x p(x|g) 
    f_g(x),\\
    \nonumber &=- \log 
    \sum_g \max_{\{p(x|g)\}} \sum_x p(x|g) {\rm Tr}[M_g \Lambda_x(\rho)]p(x),\\
    \nonumber &=-\log \max_{\{p(x|g)\}} \sum_x {\rm Tr}\left[\left( \sum_g p(x|g) M_g \right) \Lambda_x(\rho) \right]p(x),\\
    \nonumber &= -\log  
    \max_{\mathbb{N}\prec \mathbb{M}} \sum_x {\rm Tr}[N_x \Lambda_x(\rho)]p(x)
    ,\\
    &= -\log P^{\rm D}_{\rm succ}(\Lambda,\mathbb{M},\rho).
    \label{eq:b4r}
\end{align} 
We then have the following expression:
\begin{align*}
    &I_{+\infty}(X_{\Lambda,\rho} \colon G_\mathbb{M})-
    \max_{\sigma\in {\rm F}}
    \max_{\mathbb{N}\in \mathbb{F}}
    I_{+\infty}(X_{\Lambda,\sigma} \colon G_\mathbb{N})\\
    &=-H_{+\infty}(X_{\Lambda,\rho}|G_\mathbb{M})-
    \max_{\sigma\in {\rm F}}
    \max_{\mathbb{N}\in \mathbb{F}}
    -H_{+\infty}(X_{\Lambda,\sigma}|G_\mathbb{N}),\\
    &=\log 
    \Big [
    P^{\rm D}_{\rm succ}(\Lambda,\mathbb{M},\rho)
    \Big ]
    - 
    \max_{\sigma\in {\rm F}}
    \max_{\mathbb{N}\in \mathbb{F}}
    \log \Big [
    P^{\rm D}_{\rm succ}(\Lambda,\mathbb{N},\sigma)
    \Big ],\\
    &= \log \left \{
    \frac{P^{\rm D}_{\rm succ}(\Lambda,\mathbb{M},\rho)
    }{
    \max_{\mathbb{N}\in \mathbb{F}}
    \max_{\sigma\in {\rm F}}
    P^{\rm D}_{\rm succ}(\Lambda,\mathbb{N}, \sigma)
    }
    \right \}.
\end{align*}
We now maximise over all ensembles of channels and using Result 2A we obtain the claim in \eqref{eq:result3RA}.
\end{proof}

\textbf{Result 3B:} The maximum gap between the order minus-infinity mutual information of \emph{any} state-measurement pair $(\rho,\mathbb{M})$ when compared to the best \emph{fully free} state-measurement pair is upper bounded as:
\begin{align}
    \nonumber &\max_{\Lambda} 
    \left\{
    I_{-\infty}(X_{\Lambda,\rho} \colon G_{\mathbb{M}})
    -
    \max_{\sigma \in {\rm F}}
    \max_{\mathbb{N} \in \mathbb{F}}
    I_{-\infty}(X_{\Lambda,\sigma} \colon G_{\mathbb{N}})
    \right\}\\
    &\leq
    -\log \Big[1-{\rm W_F}(\rho)\Big]
    -\log \Big[1-{\rm W_\mathbb{F}}(\mathbb{M})\Big],
    \label{eq:result3WA}
\end{align}
with the maximisation over all ensembles of channels.

\begin{proof}
The minus-infinity mutual information between classical random variables $X_{\Lambda,\rho}$ and $G_\mathbb{M}$ is given by \cite{QRT1, QRT2}:
\begin{align*}
    I_{-\infty}(X_{\Lambda,\rho} \colon G_\mathbb{M})=-
    \left[
    H_{-\infty}(X_{\Lambda,\rho}|G_\mathbb{M})-H_{-\infty}(X_\Psi)
    \right],
\end{align*}
with $H_{-\infty}(X_{\Lambda,\rho})=- \log (\min_x p(x))$, $H_{-\infty}(X_{\Lambda,\rho}|G_\mathbb{M})=-\log \sum_g \min_x p(g,x)$, $p(g,x)=p(g|x)p(x)$. Using $p(g|x)=\Tr[M_g\Lambda_x(\rho)]$ then $H_{-\infty}(X_{\Lambda,\rho}|G_\mathbb{M})=-\log \sum_g \min_x \Tr[M_g\Lambda_x(\rho)]p(x)$. Considering $f_g(x)={\rm Tr}[M_g \Lambda_x(\rho)]p(x)$ and using:
\begin{align}
    \min_x f_g(x) =\min_{\{p(x|g)\}} \sum_x p(x|g) f_g(x),
    \label{eq:p2}
\end{align}
we have:
\begin{align} 
    & H_{-\infty}(X_{\Lambda,\rho}|G_\mathbb{M})
    \nonumber\\
    &=- \log 
    \sum_g \min_{\{p(x|g)\}} \sum_x p(x|g) 
    f_g(x)
    , \nonumber\\
    \nonumber &=- \log 
    \sum_g \min_{\{p(x|g)\}} \sum_x p(x|g) {\rm Tr}[M_g \Lambda_x(\rho)]p(x),\\
    \nonumber &=-\log \min_{\{p(x|g)\}} \sum_x {\rm Tr}\left[\left( \sum_g p(x|g) M_g \right) \Lambda_x(\rho) \right]p(x),\\
    \nonumber &= -\log  
    \min_{\mathbb{N}\prec\mathbb{M}} \sum_x {\rm Tr}[N_x \Lambda_x(\rho)]p(x)
    ,\\
    &= -\log P^{\rm E}_{\rm err}(\Lambda,\mathbb{M},\rho).
    \label{eq:b4w}
\end{align}
We then have the following expression:
\begin{align*}
    &I_{-\infty}(X_{\Lambda,\rho}|G_\mathbb{M})
    -
    \max_{\sigma\in {\rm F}}
    \max_{\mathbb{N}\in \mathbb{F}}
    I_{-\infty}(X_{\Lambda,\rho}|G_\mathbb{N})\\
    &=H_{-\infty}(X_{\Lambda,\rho}|G_\mathbb{M})-
    \max_{\sigma\in {\rm F}}
    \max_{\mathbb{N}\in \mathbb{F}}
    H_{-\infty}(X_{\Lambda,\sigma}|G_\mathbb{N}),\\
    &=-\log 
    \Big [
    P^{\rm E}_{\rm err}(\Lambda,\mathbb{M},\rho)
    \Big ]- 
    \max_{\sigma\in {\rm F}}
    \max_{\mathbb{N}\in \mathbb{F}}
    -\log \Big [
    P^{\rm E}_{\rm err}(\Lambda,\mathbb{N},\sigma)
    \Big ],\\
    &
    =-\log 
    \Big [
    P^{\rm E}_{\rm err}(\Lambda,\mathbb{M},\rho)
    \Big ]+ 
    \min_{\sigma\in {\rm F}}
    \min_{\mathbb{N}\in \mathbb{F}}
    \log \Big [
    P^{\rm E}_{\rm err}(\Lambda,\mathbb{N},\sigma)
    \Big ],\\
    &=-\left \{ \log 
    \Big [
    P^{\rm E}_{\rm err}(\Lambda,\mathbb{M},\rho)
    \Big ]
    -
    \min_{\sigma\in {\rm F}}
    \min_{\mathbb{N}\in \mathbb{F}}
    \log \Big [
    P^{\rm E}_{\rm err}(\Lambda,\mathbb{N},\sigma)
    \Big ]
    \right \},\\
    &=-\log \left \{
    \frac{
    P^{\rm Q}_{\rm err}(\Psi,\mathbb{M},\rho)
    }{
    \min_{\sigma\in {\rm F}}
    \min_{\mathbb{N}\in \mathbb{F}}
    P^{\rm E}_{\rm err}(\Psi,\mathbb{N},\sigma),
    }
    \right \}.
\end{align*}
We now maximise over all ensembles of channels and using Result 2B we obtain the claim in \eqref{eq:result3WA}.
\end{proof}

\end{document}